%% file: main.tex
\PassOptionsToPackage{dvipsnames}{xcolor}

\documentclass[sigconf,10pt,screen]{acmart}
\AtBeginDocument{%
  \providecommand\BibTeX{{%
    \normalfont B\kern-0.5em{\scshape i\kern-0.25em b}\kern-0.8em\TeX}}}

\setcopyright{none}

\acmConference[]{}{}{}
\acmYear{}
\acmISBN{} 
\acmDOI{} 

\usepackage{url}
\usepackage{listings}
\usepackage{multirow}

\newcommand{\REM}[1]{}

\usepackage{fancyvrb}
\usepackage{float}
\usepackage{hyperref}
\usepackage{booktabs}
\usepackage{multirow}
\newsavebox{\codebox}
\usepackage[caption=false]{subfig}
\usepackage{backnaur}
\usepackage{xcolor}
\usepackage{pgfplots} 
\usetikzlibrary{patterns}

\usepackage{comment}

\newcommand {\todo}[1]{\textcolor{blue}{TODO: #1}}

\begin{document}

\title[UPIR: Toward the Design of Unified Parallel Intermediate Representation]{\huge UPIR: Toward the Design of Unified Parallel Intermediate Representation for Parallel Programming Models}

\author{Anjia Wang}
\email{awang15@uncc.edu}
\affiliation{%
  \institution{University of North Carolina at Charlotte}
  \city{Charlotte}
  \state{North Carolina}
  \country{USA}
  \postcode{28223}
}

\author{Xinyao Yi}
\email{xyi2@uncc.edu}
\affiliation{%
  \institution{University of North Carolina at Charlotte}
  \city{Charlotte}
  \state{North Carolina}
  \country{USA}
  \postcode{28223}
}

\author{Yonghong Yan}
\email{yyan7@uncc.edu}
\affiliation{%
  \institution{University of North Carolina at Charlotte}
  \city{Charlotte}
  \state{North Carolina}
  \country{USA}
  \postcode{28223}
}

\renewcommand{\shortauthors}{Anjia Wang, et al.}

\begin{abstract}
\input{src/abstract}
\end{abstract}


%


\maketitle

\section{Introduction}
\label{sec:intro}
\input{src/intro}

\section{Motivation and Related Work}
\label{sec:motivation}
\input{src/motivation}

\section{The UPIR for Parallelism: SPMD, Data and Asynchronous Tasking}
\label{sec:upir-parallelism}

\input{src/upir-parallelism}

\section{The UPIR for Data Attribute, Explicit Data Movement and Memory Management}
\label{sec:upir-data}
\input{src/upir-data}

\section{The UPIR for Synchronization, Communication and Mutual Exclusion}
\label{sec:upir-sync}
\input{src/upir-sync}

\section{Evaluation}
\label{sec:eva}
\input{src/eva}


\section{Conclusion}
\label{sec:conclusion}
\input{src/conclusion}


\begin{acks}
This material is based upon work supported by the National Science Foundation under Grant No. 1833332 and 2015254.
\end{acks}

\bibliographystyle{ACM-Reference-Format}
\bibliography{references}

\end{document}

%% file: src/abstract.tex
The complexity of heterogeneous computing architectures, as well as the demand for productive and portable parallel application development, have driven the evolution of parallel programming models to become more comprehensive and complex than before. 
Enhancing the conventional compilation technologies and software infrastructure to be parallelism-aware has become one of the main goals of recent compiler development. 
In this paper, we propose the design of unified parallel intermediate representation (UPIR) for multiple parallel programming models and for enabling unified compiler transformation for the models. UPIR specifies three commonly used parallelism patterns (SPMD, data and task parallelism), data attributes and explicit data movement and memory management, and synchronization operations used in parallel programming. We demonstrate UPIR via a prototype implementation in the ROSE compiler for unifying IR for both OpenMP and OpenACC and in both C/C++ and Fortran, for unifying the transformation that lowers both OpenMP and OpenACC code to LLVM runtime, and for exporting UPIR to LLVM MLIR dialect. 



%% file: src/intro.tex
The past two decades have seen dramatically increased complexity of computer systems, including the significant increase of parallelism from 10s to 100s and 1000s computing units and cores, the wide adoption of heterogeneous architecture such as CPU, GPUs and vector units in a computer system, and the significant enhancement to the conventional memory hierarchy using new memory technologies such as 3D-stacked memory and NVRAM. Demands from users and applications for computing have also become high and diverse, ranging from computational science, large-scale data analysis, and artificial intelligence that adopts the computation-intensive deep neural network methods. Together they have driven the evolution of parallel programming models to become more comprehensive and complex with multifaceted goals including delivering portable performance across diverse architectures, being highly expressible for the wide ranges of users and applications, and allowing for high performance implementation and tools support.  

Compilers have been playing the critical role to meet those goals for parallel programming. 
Enhancing the conventional compilation technologies and software infrastructure to be parallelism-aware has become one of the main goals of recent compiler development. 
However, despite the efforts to support parallelism-aware compilation in existing compilers~\cite{kotsifakou2018hpvm,sharif2019approxhpvm,majeti2015heterogeneous, doerfert2019tregion,tiotto2019openmp,prabhu2020chunking} and efforts to augment compiler intermediate representation (IR) with parallelism~\cite{schardl2017tapir,maydan1993array,chen2018tvm,ragan2013halide,benoit2010extending,benoit2011kimble,lattner2021mlir}, compilers may still generate sub-optimal parallel code~\cite{georgakoudis2020faros}. 
It is also observed that existing parallel programming models share common parallelism functionality and use similar interfaces of essential capability for programming parallelism~\cite{salehian2017comparison}. However, supporting these parallel models  in one compiler often has to create language-dependent compiler passes of the same functionality for different models.
We believe one of the barriers is the lack of language-independent abstraction of the fundamental entities and constructs for parallelism. This has hindered the research and development of parallelism-aware analysis and transformation across multiple programming models.

In this paper, we propose the notion and specification of {\it unified parallel intermediate representation (UPIR)} to enable language-neutral parallelism-aware compilation.
UPIR specifies 1) three commonly used parallelism patterns, namely single program multiple data (SPMD), data parallelism, and task parallelism including offloading tasks; 2) data attributes and explicit data movement and memory management for assisting data-aware optimization for parallel programs; and 3) synchronization operations (e.g. barrier, reduction, mutual exclusion, etc) used in parallel programming for optimizing synchronization cost by the compiler. We create a prototype implementation in ROSE compiler, and demonstrate UPIR for unifying IR for offloading code in both OpenMP and OpenACC and in both C/C++ and Fortran. The demonstration also includes a unified transformation that lowers both OpenMP and OpenACC offloading code to LLVM OpenMP runtime. UPIR is also implemented as LLVM MLIR dialect, thus the ROSE-based UPIR compiler is able to export the UPIR of a program to its MLIR dialect. 

For the remainder of the paper, 
Section~\ref{sec:motivation} discusses the related work and our motivation. In 
Section~\ref{sec:upir-parallelism}, we describe the design of UPIR for three kinds of parallelism, i.e. SPMD, data and task parallelism. Section~\ref{sec:upir-data} includes the description of UPIR for specifying data attributes, data movement and memory management used in parallel programming. In Section~\ref{sec:upir-sync}, we describe the UPIR for synchronization in parallel programming. 
Section~\ref{sec:eva} includes the evaluation of the UPIR to support multiple parallel models and unified compiler transformation. Section \ref{sec:conclusion} concludes the paper.

%% file: src/motivation.tex
\REM {
\begin{table}[!ht]
\centering
\resizebox{\linewidth}{!}{
\begin{tabular}{|l|l|l|l|}
\hline
\textbf{Related Work} & \textbf{Year} & \textbf{Use Case} & \textbf{Parallellism} \\ \hline
Lee et al.\cite{lee2012efficient} & 2012 & MHP & Generic IR \\ \hline
Chen et al.\cite{chen2012can} & 2012 & MHP & Generic IR \\ \hline
Agarwal et al.\cite{agarwal2007may} & 2007 & MHP & Generic IR \\ \hline
Mellor-Crummey\cite{mellor1993compile} & 1993 & Data Race Detection & Generic IR \\ \hline
Voung et al.\cite{voung2007relay} & 2007 & Data Race Detection & Generic IR \\ \hline
Chen et al.\cite{chen2013pruning} & 2013 & Data Race Detection & Generic IR \\ \hline
Prabhu et al.\cite{prabhu2020chunking} & 2020 & Loop Chunking & Generic IR \\ \hline
Maydan et al.\cite{maydan1993array} & 1993 & Data Privatization & Parallel IR: Last Write Tree \\ \hline
Chen et al.\cite{chen2018tvm} & 2018 & DSL: deep learning & Parallel IR: TVM IR \\ \hline
Ragan-Kelly et al.\cite{ragan2013halide} & 2013 & DSL: image processing & Parallel IR: Halide IR \\ \hline
Tapir\cite{schardl2017tapir,schardl2019tapir} & 2017 & Generic & Parallel IR: Tapir \\ \hline
Kotsifakou et al.\cite{kotsifakou2018hpvm} & 2018 & Generic & Parallel IR: HPVM \\ \hline
Sharif et al.\cite{sharif2019approxhpvm} & 2019 & Generic & Parallel IR: approxHPVM \\ \hline
Benoit et al.\cite{benoit2010extending} & 2010 & Generic & Parallel IR: Gomet \\ \hline
Benoit et al.\cite{benoit2011kimble} & 2011 & Generic & Parallel IR: KIMBLE \\ \hline
Khaldi et al.\cite{khaldi2012spire} & 2012 & Generic & Parallel IR: SPIRE \\ \hline
Jordan et al.\cite{jordan2013inspire} & 2013 & Generic & Parallel IR: INSPIRE \\ \hline
Yan et al.\cite{yan2009hierarchical} & 2009 & Generic & Generic IR \\ \hline
Majeti et al.\cite{majeti2015heterogeneous} & 2015 & Generic & Generic IR \\ \hline
Doerfert et al.\cite{doerfert2019tregion} & 2019 & Generic & Generic IR \\ \hline
Tiotto et al.\cite{tiotto2019openmp} & 2019 & Generic & Generic IR \\ \hline
\end{tabular}
}
\caption{Studies on parallel IR/compilation}
\label{tab:parallelism_relate_work}
\end{table}

}


Many compiler implementations to support the compilation of parallel programs take the approach of transforming (lowering) IRs of parallel constructs to API calls of parallel runtime systems, and it is often performed at the compiler front-end. The parallel programs are then considered as sequential programs in the later stages of compilation. There are, in general, at least two limitations of this approach.
First, the parallelism information may not be carried sufficiently through the whole compilation pass to the back-end, and the compilation would then lose some optimization opportunities in the later stage~\cite{schardl2017tapir}. 
For example, function outlining is a common compiler transformation step for lowering OpenMP {\it parallel} code region to runtime API calls.
The new outlined function is separated from the original function. Thus traditional data- and control-flow analysis and optimization cannot apply to the original function body without additional effort to associate the two functions after outlining. 
Secondly, since the transformation for parallelism is performed at the front-end, operating on language-specific IRs, parallelism-aware compilation becomes language-specific. For each programming model, it requires implementing the transformation in the compiler front-end.
The study of thread programming models such as OpenMP, OpenACC, CUDA, Cilk, etc. shows that programming models share common functionality and even similar syntax for programming parallelism~\cite{salehian2017comparison}. Thus it is possible to unify compiler transformation for multiple models if a common and language-neutral IR is used by the compiler~\cite{jordan2013inspire}. Major open source compilers such as GCC and LLVM have shown unification of IRs for multiple languages, their support for unifying parallel models are however limited. 
\REM{

Second, compilers may not provide enough parallelism information due to imperfect parallel IR design, and it causes less accurate analysis.
For instance, the compile-time static data race detection could trigger excessive false positives \cite{chen2013pruning}.
Lastly, existing optimizations may only target specific languages or compilers.  
Doerfert et al. proposed a new interface \texttt{TRegion} used in the OpenMP lowering procedure \cite{doerfert2019tregion}.
It postpones the decision of target-specific transformation and finds a better solution rather than solely relying on the user's original OpenMP source code.
Tiotto et al. presented another optimization for OpenMP GPU offloading \cite{tiotto2019openmp}.
They used static analysis to detected the best grid size of \texttt{teams} construct on the IBM XL compiler.
To migrate these optimizations to OpenACC or LLVM, developers have to duplicate the implementation.

, and 2) performing optimizations as if it is a sequential program. 

Traditional compiler transformations are designed for serial programs.
They may fail to perform analysis, transformation, or optimization on parallel programs.
Parallelism-aware compilation addresses this issue by carrying the parallelism information, and it is beneficial in many cases.
May-Happen-in-Parallel analysis determines the behaviors of a parallel program, specifically, whether a given pair of statements or a single statement can be executed in parallel \cite{lee2012efficient,chen2012can,agarwal2007may}.
Besides the uncertain performance and behavior, parallel programs may have incorrect execution results caused by data race.
The compile-time static analysis could find data race and reduce the amount of probing operations at runtime \cite{mellor1993compile,voung2007relay}.
Prabhu et al. proposed generating loop chunks based on the estimation of loop iterations' computing intensity at compile-time, instead of using one fixed chunk size \cite{prabhu2020chunking}.



\subsubsection{MPI communication}
Compiler optimization for MPI communication can be roughly divided into two parts: 1)MPI communication and computation overlapping and 2)MPI communication data type checking.
Considering that traditional compilers explore no details of the calls to the MPI library, Danalis et.al~\cite{danalis2009mpi} presents an algorithm to optimizing compiler transformation to improve the overlap of MPI communication and computation.
The side effects of key MPI functions are marked and archived solution for corresponding effect can be applied according to the mark.
Guo et.al~\cite{guo2016compiler} presents a method to improve MPI communication and computation overlap by using flow graph.
Once the hot spots or potential hot spots of MPI communication are identified, sufficient surrounding loops are collected from flow graph to overlap MPI communications.
Das et.al~\cite{das2008compiler} presents an aggressive optimizing compiler to extract opportunities of overlapping MPI communication and computation automatically.
After that, for MPI applications, \texttt{mpi\_wait} can be removed and \texttt{mpi\_send/mpi\_recv}can be split to create more overlap.

Not only reducing execution time, lowering memory overhead is also a way to do optimization.
TypeART~\cite{huck2018compiler} works as an MPI correctness checker to track memory allocations and deallocations related to MPI communication.
Since it is developed from MUST~\cite{hilbrich2013mpi}, which is a runtime correctness checker for MPI, after extracting type information and inserting instrumentation to track memory allocations on compiler-time, the tracking work is actually done by MUST on runtime.
MPI-CHECK~\cite{luecke2003mpi} covers messages checking including insufficient space and negative length.
While, only Fortran 90 and Fortran 77 are allowed in this case.

However, these implementations are language specific.
They concentrate on MPI only and it is complicated to extent the usage to other parallelism programming language such as OpenMP.
We realize that for IR based implementations, the compile-time data distribution scheduling is quite the same thing among various parallelism programming languages, then we present a language neutral implementation to make it wilder adaptable comparing with the language specific ones.

As a summary, existing work are ad-hoc approach and there is no standard abstraction for parallel program structure for assisting those analysis and transformation. There is no unified and comprehensive approach and abstraction for representing parallel flow graph.

\subsubsection{May-Happen-in-Parallel Analysis (mostly for async tasking)}

Parallel programming could be very complex, and sometimes its execution result is difficult to predict due to the nature of parallelism.
Thus, may-happen-in-parallel (MHP) analysis is conducted to determine the behaviors of a parallel program, specifically, whether a given pair of statements or a single statement can be executed in parallel.
There have been a variety of MHP algorithms and corresponding implementations \cite{lee2012efficient,chen2012can,agarwal2007may}.
Agarwal et al. introduced the program structure tree (PST) to perform never-executed-in-parallel (NEP) analysis and place-equivalent (PE) analysis. Then the results are combined to obtain MHP information.
They claimed that it's a more efficient and accurate approach since the based language X10 has better concurrency constructs, which is favorable for more precise analysis.
However, the fact that the study depends on the properties of a specific language makes it difficult to be ported to other languages.

\subsubsection{Data privatization or other related data-flow analysis for parallel programs}

In traditional serial program analysis, access to an array is considered as that the whole array is accessed.
However, in parallel programs, this assumption could trigger false dependencies.
Multiple threads may access different parts of an array without any overlapping.
For a scalar variable, the dependency in continuous read and write operations could be eliminated by renaming if the operations to the same variable can be split into multiple isolated stages.

\subsubsection{Data Race Detection}

Besides the uncertain performance and behavior, parallel programs may have incorrect execution results caused by data race.
Data race happens when multiple threads try to access the same data, and at least one writes to the data.
Due to the unpredictable property of parallel programs, it's preferred to detect data race in static reliably before executing the program.
Mellor-Crummey presented ERASER to conduct compile-time analysis to find data race and reduce the amount of probing operations for race detection at runtime significantly \cite{mellor1993compile}.
Vound et al. developed a static data race detection tool Relay that can be scaled to millions of lines of code \cite{voung2007relay}. It builds a call graph and determines the lockset for memory accesses.
However, due to lacking parallelism awareness, Relay may provide too many false positives.
Chen et al. point out that Relay fails to recognize the serial section,  barrier, and conditional execution in a parallel region \cite{chen2013pruning}.
To resolve this issue, they perform code region analysis and phase analysis to determine single-threaded regions, parallel regions, non-parallel regions, and barriers inside parallel regions.

\subsubsection{Loop Chunking}

To speed up a loop execution, its iterations can be split into multiple chunks and handled by different threads running in parallel.
Regardless of the scheduling policy, the size of chunks is primarily even.
However, depending on the loop body, each iteration may have a significantly different workload so that each chunk could cost different execution times \cite{prabhu2020chunking}.
Therefore, Prabhu et al. proposed deep chunking to generate chunks based on the workload.
It analyzes the serial and parallel regions to estimate the time cost at compile-time.

\subsubsection{Vectorization}
IR for vectorization is more just affine loop representation. Matrix or polygon representation is widely used if polyhedral transformations are applied. Focus on some related work on polyhedral analysis for vectorization.

}

\subsection{GCC}

GCC 
supports a wide range of programming languages and platforms.
It uses a language independent IR GIMPLE.
GCC implements different parser for each language and generates a generic AST first.
Then, it converts the generic AST into GIMPLE for further analysis and transformation.
This design reduces the duplicated work of implementing similar compilation passes for multiple languages.
GCC supports both OpenMP and OpenACC.
A set of IR components are added as extension to GIMPLE for OpenMP, such as GIMPLE\_OMP\_TARGET representing an offloading region.
All OpenMP constructs are converted to corresponding GIMPLE IR objects.
For OpenACC, GCC maps OpenACC constructs to OpenMP GIMPLE IRs.
For instance, \textit{acc kernels} and \textit{acc parallel} are both converted to GIMPLE\_OMP\_TARGET.
The transformation for OpenACC then does not need to be implemented separately. 

The unification for OpenMP and OpenACC supports in GCC proved that a unified parallel IR reduces the compiler programming effort and an existing transformation module can be extended easily to support another parallel model.
However, rather than working as an abstraction for multiple parallel programming models, those GIMPLE IR objects are primarily designed for OpenMP and tightly associated with OpenMP syntax and semantics.
Not all the necessary parallelism information is included since they may not be required according to OpenMP specification.
Given a group of optimization passes that need complete data attributes, each of them have to search individually because GIMPLE only contains the information specified explicitly by users.

\subsection{LLVM}

LLVM is another widely used compiler like GCC, but with a more flexible modular design.
Different languages have their own front-ends that generate unified LLVM IR for later transformation. 
LLVM IR is a low-level language-neutral IR to support the same set of compiler analysis and transformation for multiple programming models. 
However, the LLVM IR was not designed for parallelism, though it can be enhanced to encode certain parallelism information using metadata or IR attribute (e.g. the {\it llvm.loop.parallel\_accesses}~\cite{llvm_compiler:web}), or parallel intrinsic functions~\cite{tian2016llvm}. 
However, using the intrinsic or IR metadata has the same limitation of being complicated when the parallel construct becomes complicated. Also, the conventional serial analysis and optimization may not be applicable anymore because the parallel region has been outlined and the compiler is not aware of the semantics of intrinsic calling the outlined function. To bridge the gap between the low-level LLVM IR and high-level language-specific IR and frontend, recently, LLVM adopts the MLIR~\cite{lattner2021mlir} as the middle-end IR and abstraction to address the fragmentation of compiler development for multiple programming models including domain specific languages.
However, until now, MLIR has mainly been used for defining language-specific IRs dialect for new languages or extension to existing languages, but not for unifying IRs of common constructs of multiple languages \cite{mlir_dialects:web}. 
\subsection{Enhancing Existing IR for Parallelism}

\REM{
However, using metadata for parallelism might not flexible enough when parallel programming constructs becomes complex. 
For example, \textit{llvm.loop.parallel\_accesses} in LLVM denotes the memory access in a parallelizable loop \cite{llvm_compiler:web}.
Nevertheless, the compiler cannot optimize the loop freely because inserting any instructions without this metadata will break the parallelism.

\subsubsection{Intrinsic functions} 

Parallelism notations in a program are transformed to API functions, such as \textit{\_\_kmpc\_fork\_call} in LLVM~\cite{llvm_compiler:web} and \textit{GOMP\_parallel\_start} in GCC \cite{gnu_compiler:web}. There are efforts in developing parallel intrinsic functions in LLVM to support OpenMP~\cite{tian2016llvm} with LLVM IR. However, using the intrinsic functions has the same limitation of being complicated when the parallel construct become complicated. 
Also the conventional serial analysis and optimization may not be applicable anymore because the internal implementation and dependencies are transparent to the compiler.
}


There are proposals of new IRs for parallelism, including as much relevant information as possible.
Zhao et al. designed a set of parallel IR including high, middle, and low levels to represent parallel constructs \cite{zhao2011intermediate}, focusing on parallel loop chunking, load elimination, delegated isolation, automatic data privatization, vectorization, and other cases.
Benoit et al. presented Gomet and KIMBLE as GCC extensions to support parallel IR \cite{benoit2010extending,benoit2011kimble}.
It introduces a hardware abstraction and hierarchy of parallel IR for parallelism adaption. 
Doerfert et al. proposed a new interface \texttt{TRegion} used in the OpenMP lowering procedure \cite{doerfert2019tregion}.
It postpones the decision of target-specific transformation and finds a better solution rather than solely relying on the user's original OpenMP code.
HPVM/approxHPVM is another design of parallel IR~\cite{kotsifakou2018hpvm,sharif2019approxhpvm}.
HPVM provides compiler IR, virtual ISA, and runtime libraries, while other studies mainly focus on one of them. Tapir/LLVM~\cite{schardl2017tapir} adds three instructions \textit{detach}, \textit{reattach}, and \textit{sync} to encode logical parallelism asymmetrically in LLVM IR to support task parallelism. It has been extended to support OpenMP tasking~\cite{stelle2017openmpir} and TensorFlow~\cite{schardl2019tapirxla}. However, these state of the art lack the generality for supporting multiple parallelism patterns (e.g. SPMD, data and task parallelism), data sharing attributes and synchronization operations of parallel program. 

\subsection{Creating Unified Parallel Intermediate Representation (UPIR)}

As existing parallel programming models share common parallelism functionality and similar interfaces of essential capability for programming parallelism~\cite{salehian2017comparison}, a language-independent abstraction of the fundamental entities and constructs for parallelism and their connections can be constructed in a unified intermediate representation as the backbone to enable unified and common parallelism-aware analysis and transformation. The UPIR design and specification include  1) the three commonly used parallelism patterns, namely single program multiple data (SPMD), data parallelism, and task parallelism including offloading tasks; 2) data attributes and explicit data movement and memory management for assisting data-aware optimization for parallel programs; and 3) synchronization operations (e.g. barrier, reduction, mutual exclusion, etc) used in parallel programming for optimizing synchronization cost by the compiler. Table~\ref{tab:parallelism_summary} shows the UPIP's support and mapping for the language constructs of commonly-used parallel models. 
In the following three sections, we present our design of the UPIR. 
Since UPIR naturally unifies IR of different programming models, an additional benefit is that supporting programs of hybrid models or supporting translating program of one model to another would be made much easier. 
In comparison with related work such as INSPIRE~\cite{jordan2013inspire}, they share a similar concept, but UPIR’s design allows for more parallelism information in the IR and provides more features than INSPIRE. For example, UPIR categorizes data parallelism into worksharing, SIMD, and taskloop, compared to a generic work distribution in INSPIRE.

\subsubsection{UPIR extension}

Since the goal of UPIR is to cover the the common features shared by multiple parallel models, thus the limitation would be that some of the unique features of a parallel model might not be included in UPIR. e.g., memory controls in CUDA. In our design, we address this by including UPIR extension in the form key-value maps such that language-specific features can be added under the UPIR. For compiler, language-unique compiler pass can be added as extension of the UPIR transformation pass in the UPIR compiler.
For instance, \textit{metadirective} in OpenMP is created as an UPIR extension but not part of the UPIR nodes. It will be handled by the compiler only if a corresponding transformation pass is implemented.



\begin{table*}[!htb]
\centering
\resizebox{\textwidth}{!}{%
\begin{tabular}{|l|l|l|l|l|l|}
\hline
 & \textbf{SPMD parallelism} & \textbf{Data parallelism} & \textbf{Async task parallelism} & \textbf{Data attributes} & \textbf{Synchronization} \\ \hline
UPIR & spmd & loop/loop-parallel & task & data & sync \\ \hline
CUDA & \textless{}\textless{}\textless{}...\textgreater{}\textgreater{}\textgreater{} & - & async kernel launching & \_\_shared\_\_, \_\_global\_\_, memory operations & cudaDeviceSynchronize \\ \hline
OpenCL & clEnqueueNDRangeKernel & - & clEnqueueTask & clCreate/Read/WriteBuffer & clWaitForEvents \\ \hline
OpenClik & - & cilk\_for, array operations, elemental functions & cilk\_spawn, cilk\_sync & - & cilk\_sync \\ \hline
Kokkos & {\bf parallel}\_for & parallel\_{\bf for} & task\_spawn, host\_spawn & View, memory space & fence \\ \hline
SYCL & {\bf parallel}\_for & parallel\_{\bf for} & single\_task, host\_task & buffer & wait \\ \hline
OpenMP & teams, parallel & distribute, for, simd & task, target & map(to/from), shared/private/firstprivate & barrier, atomic, critical, taskwait \\ \hline
OpenACC & parallel & loop gang/worker/vector & async, wait & data(copyin/out), shared/private/firstprivate & wait, atomic \\ \hline
PThread & pthread\_create/join & - & - & - & pthread\_join, pthread\_cond\_wait \\ \hline
\end{tabular}%
}
\caption{Mapping of parallel model constructs with UPIR design}
\label{tab:parallelism_summary}
\end{table*}



\REM{
\subsubsection{Encode parallelism using existing IR}

Instead of creating new IR, an alternative way to handle parallel programs is to use existing IR to represent parallelism.
Compilers do not need to change their existing optimization and transformation heavily.
Tapir/LLVM adopts the approach  to insert a pass for parallelism before converting the source code to LLVM IR \cite{schardl2017tapir,schardl2019tapir, stelle2017openmpir}.
Tapir adds three instructions \texttt{detach}, \texttt{reattach}, and \texttt{sync} to encode logical parallelism asymmetrically in a serial AST to support task parallelism.
Then any serial analysis can be performed since three new instructions will be lowered to the existing IR.
However, Tapir is targeting compilation solely but does not consider other use cases much, like data race detection.
This approach is not considered a perfect solution due to lacking parallelism awareness \cite{doerfert2018compiler}.


\subsection{Motivation}



By investigating the states of the arts for parallel program compilation, we noticed that the existing solutions 1) may not work as compiler IR and benefit from traditional optimization passes, 2) lack some parallel information as parallel IR, 3) are not portable across multiple platforms or toolchains,
4) are designed particularly for one parallel programming model but not applicable to others.
Therefore, we propose a unified parallel intermediate representative (UPIR), which includes as much parallelism information as possible, provides one unified compiler transformation for multiple parallel programming languages,  and can be exported as language-neutral MLIR.
}

%% file: src/upir-parallelism.tex
\REM{
\begin{table*}[!ht]
\resizebox{\textwidth}{!}{
\begin{tabular}{|l|l|l|}
\hline
\textbf{Architecture and application} & \textbf{Analysis/Optimization/Transformation}   & \textbf{IR}                                                         \\ \hline
GPUs                                 & GPU data mapping                                & SPMD, task, data            \\ \hline
GPUs                                 & GPU pipelining                                  & SPMD, SPMD-branch, task, data \\ \hline
GPUs                                 & Warp divergence                                 & SPMD, SPMD-branch                                \\ \hline
CPU and GPU                          & Nested parallelism mapping                      & SPMD, nested-level                                        \\ \hline
CPU and GPU                          & Nested parallelism loop collapse                & SPMD, loop, collapse, nested-level           \\ \hline
CPU and GPU                          & Load balancing (e.g. chunk size of worksharing) & SPMD, SPMD-branch, loop                          \\ \hline
CPU                                  & Tasking (Task granularity, MHIP)                & task                                           \\ \hline
CPU                                  & taskloop                                        & task, loop, data                \\ \hline
CPU and GPU                          & Implicit data mapping                           & data                          \\ \hline
CPU and GPU                          & Redundant synchronization                        & loop, sync                                                 \\ \hline
CPU and GPU                          & Reduction                                       & SPMD, sync                                                  \\ \hline
\end{tabular}
}
\caption{Compilation use cases of UPIR}
\label{tab:pfg_mlir_use_cases}
\end{table*}
}



A programming model provides API for specifying different kinds of parallelism that either maps to parallel
architectures or facilitates expression of parallel algorithms. We
consider three commonly used parallelism patterns in parallel computing: single program multiple data (SPMD), data and asynchronous tasking. In the specification, we adopt the format of MILR dialect and specified using the EBNF form used for MILR dialect.  

\subsection{SPMD Parallelism}
Single program multiple data (SPMD) has been one of the most common styles of parallel programming that a program uses to start parallel execution. A program starts its parallel execution by launching multiple threads or processes on the processing units (hardware threads, cores, processors or nodes), and they all execute the same program  or code region (single program). During the parallel execution, each processing unit works on parts of the data (multiple data) of the program. Multiple data processing can be implemented via either programming manually, e.g. the domain decomposition method used in MPI, or via explicit data parallelism constructs (see Section~\ref{sec:upir-data}). 
Examples for the style and language constructs for SPMD include OpenMP {\it parallel} constructs, most MPI programs, 
manycore programming with GPU such as NVIDIA CUDA kernels or OpenCL kernels. 
For parallelism-aware analysis and optimization, data-race detection~\cite{mellor1993compile, chatarasi2016extended}, optimization of synchronization within a SPMD region~\cite{tseng1995compiler,oboyle1995synchronization}, data scoping and privatization~\cite{lin2004automatic}, parallel divergence analysis and reduction (e.g. NVIDIA CUDA warp divergence)~\cite{han2011reducing}, etc, are typical techniques for improving the accuracy of data race detection and performance of SPMD regions and programs. 

The IR for the SPMD parallelism model
includes the notation for SPMD and a code region. To support mapping of SPMD region to the hardware threading hierarchy such as GPUs and for languages that allows for specifying thread of hierarchy, SPMD units are organized in two-level hierarchy, namely teams and units. To create a unified IR for different usage and variation of SPMD regions or programs, other important details must be included in the IR specification for the compiler, including for example, the target parallel systems (CPU, GPU or multi-node cluster); the synchronizations used inside and at the end of the region such as barrier or reduction, and the data environment of the region, etc. 
The SPMD IR in our design is shown in Figure~\ref{fig:pfg_mlir_spmd}, which is written in EBNF form. 
The design enhances it with several important features to cover the common usage of SPMD across different programming models and to enable more advanced analysis and optimization for SPMD code regions. 

\begin{figure}[ht!]
\begin{lstlisting}[frame=single]
upir.spmd ::= 'spmd' spmd-field-list
spmd-field-list ::= spmd-field | spmd-field spmd-field-list
spmd-field ::= target | num_teams | num_units | data | nested-parent | nested-child | nested-level | branch | sync
target ::= 'target' '(' target-list ')'
target-list ::= target-item | target-item ',' target-list
target-item ::= 'cpu' | 'gpu' | 'cluster'
num_teams ::= 'num_teams' '(' expr-int ')'
num_units ::= 'num_units' '(' expr-int ')'
data ::= 'data' '(' data-list ')'
data-list ::= data-item | data-item ',' data-list
nested-level ::= 'nested-level' '(' expr-int ')'
nested-parent ::= 'nested-parent' '(' expr-id ')'
nested-child ::= 'nested-child' '(' expr-id ')'
branch ::= 'branch' '(' expr-id-list+ ')'
sync ::= 'sync' '(' expr-id-list ')'
\end{lstlisting}
\caption{UPIR MLIR dialect for SPMD parallelism specified using EBNF form}
\label{fig:pfg_mlir_spmd}
\end{figure}

\REM{
\subsubsection{SPMD Divergence}
SPMD divergence refers to the situation that the execution of the SPMD region by the participating processing units might exhibit different branches of the SPMD region. This divergence is often prescribed by programmers, such as using a mask or if-else control statement.
In this case, although all the units execute the same SPMD program or code region, they would run a different portion of the code region. Thus, the divergence might introduce load-imbalance across processing units, which is often shown as high synchronization overhead among multiple threads/processes.
For example, in the GPU execution model, this is often referred to as warp divergence and could cause significant performance loss~\cite{coutinho2011divergence}. 
To support compiler analysis and optimization for SPMD divergence, 
the SPMD IR includes a field named {\it branch} to refer to the IR regions that might contain divergences. 
}

\subsubsection{Specification for Data and Memory Management}
The SPMD IR includes the specification, using the {\it data} field, for the attributes of data and memory used by the SPMD region. For example,  the \textit{data} field can be used to specify whether a variable or an array is shared or private across processing units, which could correspond to the {\it shared} and {\it private} clauses of OpenMP for {\it parallel} directive. Our UPIR design for data specification includes fields and parameters for specifying more detailed information about data access, e.g. read-only, read-write access, data distribution and mapping, and memory management. This facilitates more advanced compiler analysis and optimization that involve data sharing and movement, such as data race detection and enabling overlapping between data movement and computations~\cite{baskaran2008automatic}. The data specification is explained in details in Section~\ref{sec:upir-data}.

\REM{
\subsubsection{Nested SPMD Regions}
Modern parallel programming models support nested parallelism, such as \textit{teams}/\textit{parallel} directives in OpenMP and \textit{gang}/\textit{worker} clauses in OpenACC.
The compiler can map them to the corresponding architecture, for example to NVIDIA GPU grid and block thread hierarchy. 
In the UPIR, if users provide the information of nesting of SPMD regions explicitly, they can be specified using the \textit{nested-level}, \textit{nested-parent} and {\it nested-child} fields. The value of
\textit{nested-level} indicates at which nested level the SPMD region is.
The innermost region is at level 0. The \textit{nested-parent} and {\it nested-child} fields are used to refer to the enclosing SPMD region and the enclosed SPMD region. 
}

\subsubsection{Synchronization of an SPMD Region}
During the execution of an SPMD region, it is often that the work units interact with each other via synchronization, e.g. barrier, and communications, e.g. shuffling or broadcasting data between units. We use the term "synchronization", in short, "sync", in a broader sense to refer to such interaction. 
In Section~\ref{sec:upir-sync}, we describe the design of synchronization UPIR for various types of sync operations.
The UPIR for the SPMD region allows for including a field to point to the UPIR objects for the synchronization used in the region. 
This facilitates compiler analysis and optimization in advance of the occurrence of the actual sync operation inside the SPMD region, enabling advanced optimization on the global level in coordination with the local level where the actual sync operations are specified. For example, the compiler can fuse a reduction operation with a barrier operation, and can eliminate redundant barriers~\cite{chen1999redundant,midkiff1987compiler} used inside the SPMD region.

\REM{

\begin{figure}[ht!]
\begin{lstlisting}[frame=single]
upir.spmdbranch ::= 'spmdbranch' field-list
field-list ::= field | field field-list
field ::= data | nested-parallelism-level | nested-parallelism-list | branch | condition | mutual-exclusion
... // most clauses are the same as the straightforward SPMD IR
// when the SPMD branch will be activated
condition ::= 'condition' '(' ssa-id ')'
\end{lstlisting}
\caption{UPIR EBNF specification for SPMD-branch}
\label{fig:pfg_mlir_spmd_branch}
\end{figure}

\subsubsection{Weakness of existing approach}

\begin{figure}[ht!]
\begin{lstlisting}[frame=single]
__attribute__((const)) int goo(const int *X, int n);
void foo(const int *X, int *Y, int n) {
  #pragma omp parallel for
    for (int i = 0; i < n; i++)
      Y[i] = X[i] + goo(X, n);
}
\end{lstlisting}
\caption{Code example of missing optimization}
\label{fig:example_missing_optimization}
\end{figure}

To process the \texttt{parallel} directive in OpenMP (Fig.~\ref{fig:example_missing_optimization}), normally the compipler will outline the parallel region and then fork multiple threads to call the outlined function repeatedly via intrisics.
However, after the outlining, the parallelism information is invisible to compiler and code structure may be changed heavily.
In this case, because the function \texttt{goo} is in a parallel region and then called in a outlined function.
The compiler fails to move it outside the loop and use its return value instead.
Excessive calls to \texttt{goo} dramatically decreased the performance.

\subsubsection{MPMD execution model}

Besides the popular SPMD model, PFG is able to include the information of MPMD (multiple program multiple data) execution model (Fig.~\ref{fig:pfg_mlir_mpmd}).
MPI is sometimes supposed to be a MPMD model because its processors may execute different programs.
\todo{Add more MPI information related to MPMD and modify the IR design.}
Besides the traditional fork-join model, PFG allows users to create a new unit from scratch anywhere and attach a code block to that unit.
\texttt{data} clause works the same as in the \texttt{spmd} IR to declare data attributes and movements.

\begin{figure}[ht!]
\begin{lstlisting}[frame=single]
upir.mpmd (::mlir::mpmd::MpmdOp)

operation ::= 'mpmd' clause-list

clause-list ::= clause | clause clause-list

clause ::= fork | join | create | data | cost

fork ::= 'fork' '(' mpmd-index ')'

join ::= 'join' '(' mpmd-index ')'

create ::= 'create' '(' mpmd-index ')'

mpmd-index ::= ssa-id

data ::= 'data' '(' data-list ')'

data-list ::= data-item | data-item ',' data-list

// data clause is described in Section 3.2

cost ::= 'cost' '(' ssa-id ')'
\end{lstlisting}
\caption{UPIR EBNF specification for MPMD}
\label{fig:pfg_mlir_mpmd}
\end{figure}
}

\subsection{Data Parallelism}
Data parallelism, by which multiple processing units perform the same operations on different data items, refers to the patterns or parallel APIs of decomposing computation and data among parallel processing units. It is often programmed as parallel loops and often inside an SPMD region.  
Depending on the target architectures including CPUs or GPUs, SIMD or vector units, and multi-node clusters, parallelization of data parallel loops is often programmed differently. 

For CPUs, GPUs or multi-node architecture, the common approach of programming data parallelism is to associate (implicitly or explicitly) the data parallel loops with an SPMD region, and then use the provided language constructs to prescribe how the loop iterations should be distributed to the processing units of the SPMD region.  OpenMP worksharing-loop constructs, and OpenACC loop constructs are typical constructs for annotating a loop for data parallel execution. 
For SIMD or vector units, the "parallelization", known as vectorization of data parallel loops, is performed by the compiler. Some programming languages provide language extensions or APIs for the user to prescribe how vectorization can be done by the compiler, e.g. the SIMD directive of OpenMP.

The UPIR design for data parallelism, shown in Figure~\ref{fig:pfg_mlir_work_sharing}, includes two IRs: 1) the IR for specifying canonical loops, 2) the IR for specifying loop parallelization target and details such as schedule and chunk size.
The separation allows for more flexibility and independence for the compiler to apply different loop transformations (e.g. tiling and unrolling) and parallelization (worksharing-loop or vectorization) passes in comparison to combining them into one IR. 
The IR for canonical loop specification includes information for loop trip (induction variable, range and step), collapsible level, and its data environment and sync operations such as reduction.

\begin{figure}[ht!]
\begin{lstlisting}[frame=single]
upir.loop ::= 'loop' loop-field-list
loop-field-list ::= loop-field | loop-field loop-field-list
loop-field ::= induction-var | lowerBound | upperBound | step | data | collapse | sync
induction-var ::= 'induction' '(' expr-id ')'
...
collapse ::= 'collapse' '(' expr-int ')'
sync = ::= 'sync' '(' expr-id-list ')'

upir.loop-parallel ::= 'loop_parallel' lp-list
lp-list ::= lp-list-item | lp-list-item lp-list
lp-list-item ::= worksharing | taskloop | simd

worksharing ::= 'worksharing' '(' ws-field-list ')'
ws-field-list = schedule | distribute | schedule distribute
schedule ::= 'schedule' '(' schedule-parameter ')'
schedule-parameter ::= schedule-policy | schedule-policy ',' chunk-size
schedule-policy ::= 'static' | 'dynamic' | 'guided' | 'runtime' | 'auto'
chunk-size ::= expr-int
distribute ::= 'distribute' '(' 'teams' | 'units' | 'teams,units' ')'
simd ::= 'simd' '(' simdlen ')'
simdlen ::= 'simdlen' '(' expr-int ')'
taskloop ::= 'taskloop' '(' taskloop-field-list ')'
taskloop-field-list ::= grainsize | num_tasks | grainsize num_tasks
grainsize ::= 'grainsize' '(' expr-int ')'
num_tasks ::= 'num_tasks' '(' expr-int ')'
\end{lstlisting}
\caption{UPIR MLIR dialect for data parallelism}
\label{fig:pfg_mlir_work_sharing}
\end{figure}

The IR for loop parallelization specifies three options of parallelizing canonical loops: worksharing, SIMD, and taskloop. The worksharing parallelization, more specifically, SPMD worksharing, is specified with information such as schedule policy (static, dynamic, guided, etc), chunk size and the distribution target of the SPMD region (teams or units or both). This is similar to the OpenMP standard as OpenMP includes a comprehensive list of options on how a canonical loop can be scheduled. Worksharing-annotated loops must be within an SPMD region. For SIMD parallelization, the IR includes fields and parameters such as simdlen. Clauses for OpenMP SIMD directives represent a rich set of options that we can cherry-pick as fields in the IR. 

The taskloop parallelization is the approach that many programming models use to parallelize a loop using an implicit parallel runtime system, typically tasking. For example, cilk\_for of OpenCilk, taskloop of OpenMP, for\_each in Rayon, Kokkos::parallel\_for, RAJA::forall, tbb::parallel\_for. Taskloop allows the runtime to be more flexible for scheduling loop iterations, in comparison to worksharing, since it does not require to be within an SPMD region and has less restriction than worksharing for the user to provide schedule details. There are two important fields for taskloop parallelization, grain size and num\_tasks that are used to control the granularity (or the number of tasks) of the taskloop. 

\REM{
The IR for specifying parallelization options allows for specifying more than one options such that 1) compiler can generate multi-variant codes to allow for the compiler or the runtime to choose the most performant one, and 2) more than one parallelizations, e.g. worksharing and then vectorization, can be applied to nested loops.
}

The current design includes the common and essential attributes among existing parallel programming models for specifying canonical loops and parallelization options that the compiler can apply. The separation of IRs for loop and parallelization allows for flexible addition of other transformations such as tiling and unrolling before or after parallel transformation of nested loops. As a foundation, more information that is specific to a programming model about loops and parallelization can be added, without comprising the generality of parallelism among programming models. 







\subsection{Asynchronous Task Parallelism}
In contrast to data parallelism which involves performing the same operations on a different part of the data, task parallelism is distinguished by running many tasks that perform different operations at the same time on the data. Being asynchronous means that a task (parent task) can spawn another task (child) and then the parent task continues its execution without waiting for the child task to complete. Starting from Cilk with C/C++ language extensions of adding {\it spawn} and {\it sync} API for creating and synchronizing asynchronous tasks on shared memory computing systems~\cite{frigo1998implementation}, task parallelism has become popular in mainstream programming models such as the {\it task} and {\it taskwait} directives of OpenMP, std::async and std::future starting from C++11. The concept of asynchronous tasking has been extended for supporting offloading computation and data movement on GPU devices, or for launching computation on remote computer nodes.   E.g. in CUDA, launching an asynchronous offloading kernel on GPU or an asynchronous memcpy operation can be considered as asynchronous tasking. For OpenMP and OpenACC, the directives used for offloading computation are considered as launching offloading tasks. Efforts for distributed tasking introduced in related work such as PaRSEC~\cite{parsecweb,DAGuE:PaRSEC:IPDPS:2011} have explored programming API and runtime systems of using remote tasking for applications including dense linear algebra or other scientific applications~\cite{danalis2015parsec}. 

Tasking parallelism allows users to express the full potential of parallelism that exists in the application and its algorithm implementation. 
The approach of compiler transformation and optimization of tasking impacts the overhead of task management as well as the policy and priority of task scheduling. 
For example, for compiler transformation to support asynchronous tasking, a task code region is often outlined as a function used in tasking management (in many tasking implementation such as LLVM and GNU OpenMP), or a re-entrant function needs to be created by the compiler that contains the task code region (Cilk adopted this approach~\cite{frigo1998implementation}). The approach of using re-entrant functions in Cilk and cactus stack~\cite{lee2010using} often incurs lower task management overhead than the outline approach, but requires more sophisticated compiler transformation. Choosing help-first or work-first work-stealing policy of tasking~\cite{guo2009work} also requires compiler to correctly transform the tasking code region. For analysis and optimization, may-happen-in-parallel analysis has been used often for asynchronous tasking~\cite{lee2012efficient,chen2012can,agarwal2007may}. The analysis gives compiler and users quantitative guidance for tuning the task granularity (thus the amount of prescribed parallelism in a program) to strike the balance between overhead of task management and load-balance for the runtime systems. 

In consideration of aforementioned techniques of compiler transformation to support tasking, the UPIR design must consider to include the required attributes in a tasking IR to support those transformation. Existing work such as Tapir~\cite{schardl2017tapir} demonstrated the successful usage of a tasking IR for compiler to generating high performance tasking code on shared memory systems.  
The design of UPIR for asynchronous tasking advances these related state of the art in at least three aspects: 1) unifying the three kinds of tasking into one IR: conventional tasking on shared memory systems, offloading tasking for accelerators, and remote tasking for distributed systems; 2) allowing the specification of data attributes of the task data environment; 3) including more fields that are used in many programming models, such as the field for task synchronization, task dependency, and target CPU, device or remote node for spawning a task. The specification of the UPIR for tasking is shown in Figure~\ref{fig:pfg_mlir_tasking_offloading}.
The tasking parallelism in UPIR has been implemented based on outlining. However, it has not been mapped to all the related constructs in the parallel models, like \textit{taskloop} directive in OpenMP. Users can also specify a task scheduling policy to guide the compiler transformation.

\begin{figure}[ht!]
\begin{lstlisting}[frame=single]
upir.task ::= 'task' task-field-list |  'task' offload task-field-list | 'task' remote task-field-list
task-field-list ::= task-field | task-field task-field-list
task-field ::= depend | data | sync | scheduling-policy
offload ::= 'offload' '(' device ':' device-id ')'
remote ::= 'remote' '(' device ':' device-id ')'
device ::= 'nvptx' | 'amd' | 'fpga' | 'host'
device-id ::= expr-int
depend ::= ... // similar to OpenMP depend clause
scheduling-policy ::= 'policy' '(' policy-item ')'
policy-item ::= 'help-first' | 'work-first'
\end{lstlisting}
\caption{UPIR MLIR dialect for tasking}
\label{fig:pfg_mlir_tasking_offloading}
\vspace{-0.3cm}
\end{figure}

\REM{

The dependencies on both data and tasks can be specified by \texttt{data} and \texttt{depend} clauses.
Additionally, the SPMD and worksharing IRs can be converted to the task IR as well \cite{schardl2017tapir,schardl2019tapir}.
The synchronization between tasks in PFG are discussed in Sect.\ref{subsec:synchronization}.
We adopt the same idea with Tapir of using asynchronous task to represent parallelism.
However, PFG contains more parallel information in the IR. Furthermore, users can choose from asynchronous tasking and conventional fork-join model in PFG based on their needs.

\subsubsection{Offloading tasks}
\label{subsec:pfgspec_offloading}
Instead of providing an individual IR for offloading, PFG integrates the relevant parallel information into task IR as well.
The offloaded computing is considered as a task on a remote device other than the host.
In this case, attribute \texttt{target} in task IR specifies the targeting device (Fig.~\ref{fig:pfg_mlir_tasking_offloading}).


\subsubsection{Remote tasks}

\subsubsection{Data/even-driven dependency in task parallelism}
The dependencies on data and other tasks can be described using \texttt{data} and \texttt{depend} attributes of task IR.
Particularly, the comparison of the data dependencies among tasks and parallel region can help static data race detection, and certain optimizations including kernel pipelining and overlapping.
A task IR without specifying any dependencies are free to relocated inside the parallel region by the compiler for optimization purpose.

\subsubsection{Using task to abstract data movement operations}
The data movement between host and device is also considered as a task, even though it has a dedicated IR, which will be thoroughly explained in Sect.~\ref{subsec:pfgspec_data_movement}.
If the task IR and data movement IR do not declare overlapped data usage, they can be safely executed asynchronously.
}

%% file: src/upir-data.tex






With respect to parallelism, data usage by parallel processing units is another dimension of complexity for parallel programming. 
Fundamentally, there are two kinds of operations that matter to the performance that compiler and users focus on optimizing: data movement (implicit such as paging or caching, or explicit such as memcpy), memory allocation and deallocation (memory management, mm in short). Most language-based programming models, such as OpenMP, OpenACC, and PGAS models provide language constructs for users to specify data usage attributes and let the runtime determine when and how the data movement and mm operations are performed. Library-based programming models such as MPI, CUDA and OpenCL provide APIs for those two operations that have to be explicitly invoked in a program. 

A comprehensive UPIR design should include IR fields for specifying both data attributes, as well as explicit data movement and mm operations to enable comprehensive compiler optimization across multiple programming models, such as memory-aware compilation~\cite{grun2000aggressive}, data-aware compilation~\cite{bala1993explicit} and compiler-guided data placement~\cite{li2015automatic}, overlapping data movement with computation with the help of compiler analysis~\cite{carter1994xdp}. 
Thus, we design UPIR to have three IR classes: data attribute, explicit data movement, and explicit memory allocation and de-allocation. 


\subsection{UPIR for Specifying Data Attributes}
For a data item such as a variable or an array (section) that is used during the parallel execution of a SPMD region, a data parallel loop or a task, the data attribute could include as many as six fields: 1) shared or private attribute if it is used in the shared memory system, 2) mapping attribute if it is used between discrete memory space, 3) attribute for access modes such as read-only, read-write and write-only, 4) memcpy attribute that is used to specify what memcpy API should be used when the data needs to be moved, 5) mm attribute (allocator and deallocator) that specifies what memory allocator and deallocator should be used when a new memory is needed for the data item, e.g. privatizing or mapping the data item, and 6) distribution attribute if an array (section) needs to be partitioned and distributed onto computing units. These data attributes of a data item are used to specify the intention of how data should be used for parallel execution. They do not specify when the data movement and memory allocation should happen. This leaves to compiler and runtime to apply optimization such as combining memory allocation for multiple data items, and scheduling data movement to achieve overlapping of computation and movement. 
The IR for data attribute is described in Figure~\ref{fig:pfg_mlir_data}.

\REM{
\begin{table*}[!ht]
\resizebox{\textwidth}{!}{
\begin{tabular}{|l|l|l|l|}
\hline
\textbf{Data Attribute}            & \textbf{Value}                                                 & \textbf{Optional Modifier}         & \textbf{Description}                                                                      \\ \hline
Data sharing                       & \{shared/private/firstprivate/lastprivate\}                    & Visibility = \{implicit/explicit\} & how data are shared between threads and whether it is defined implicitly                  \\ \hline
\multirow{2}{*}{Data mapping}      & \multirow{2}{*}{\{to/from/tofrom/allocate\}}                   & Visibility = \{implicit/explicit\} & \multirow{2}{*}{how data are mapped between devices and whether it is defined implicitly} \\ \cline{3-3}
                                   &                                                                & Mapper = user-defined-mapper       &                                                                                           \\ \hline
Data access                        & \{read-only/write-only/read-write\}                            &                                    & whether data are read and/or written                                                      \\ \hline
\multirow{3}{*}{Data distribution} & unit-id = user-defined-id                                      &                                    & which threads use the data                                                                \\ \cline{2-4} 
                                   & pattern = \{block/cyclic/linear/loop\}                         &                                    & in which pattern data are distributed between threads                                     \\ \cline{2-4} 
                                   & section = start-index:item-amount:stride                       &                                    & which portion of a given array is used                                                    \\ \hline
Data allocator                     & allocator = \{default\_mem\_alloc/large\_cap\_mem\_alloc/...\} &                                    & where data are allocated                                                                  \\ \hline
\multirow{2}{*}{Data movement}     & enter/exit (user-defined-id)                                   & mapping attributes                 & specify a data region and perform data movements only necessary                           \\ \cline{2-4} 
                                   & update (user-defined-id)                                       & \{to/from\}                        & manually trigger a data movement                                                          \\ \hline
\end{tabular}
}
\caption{Data attribute, memory allocation, and movement in UPIR}
\label{fig:pfg_data_attributes}
\end{table*}
}

\begin{figure}[ht!]
\begin{lstlisting}[frame=single]
upir.data ::= 'data' '(' data-list ')'
data-list ::= data-item | data-item ',' data-list
// four dimensions: data-mapping is different from data-sharing
data-item ::= expr-id '(' data-sharing ',' data-mapping ',' data-access ',' data-distribution-list ','  data-mm-allocator ',' data-mm-deallocator ',' data-memcpy ')'
data-sharing ::= data-sharing-property | data-sharing-property '(' visibility ')'
data-sharing-property ::= 'shared'|'private'|'firstprivate'|'lastprivate'
data-mapping ::= data-mapping-property | data-mapping-property '(' data-mapping-modifier-list ')'
data-mapping-modifier-list ::= data-mapping-modifier | data-mapping-modifier ',' data-mapping-modifier-list
data-mapping-modifier ::= visibility | data-mapper
data-mapper ::= ssa-id
data-mapping-property ::= 'to' | 'from' | 'tofrom' | 'allocate' | 'none'
visibility ::= 'implicit' | 'explicit'
data-access ::= 'read-only' | 'write-only' | 'read-write'
// three aspects to describe the data distribution
data-distribution-list ::= data-distribution-item | data-distribution-item ',' data-distribution-list
data-distribution-item ::= unit-id | pattern | data-section
unit-id ::= 'unit-id' '(' expr-id ')'
pattern ::= 'pattern' '(' pattern-item ')'
pattern-item ::= 'block' | 'cyclic' | 'linear' | 'loop'
data-section ::= 'section' '(' array-section+ ')'
array-section ::= '[' expr-id ':' expr-id ':' expr-id ']'
data-mm-allocator ::= 'allocator' '(' allocator-attr ')'
allocator-attr ::= 'default_mem_alloc' | 'large_cap_mem_alloc' | expr_id 
data-mm-deallocator ::= 'deallocator' '(' deallocator-attr ')'
deallocator-attr ::= 'default_mem_dealloc'|'large_cap_mem_dealloc'| expr_id 
data-memcpy ::= 'memcpy' '(' memcpy-attr ')'
memcpy-attr ::= expr_id //the memcpy function id
\end{lstlisting}
\caption{UPIR MLIR dialect for data attributes}
\label{fig:pfg_mlir_data}
\vspace{-0.3cm}
\end{figure}

Programming models such as OpenMP and OpenACC provide language construct for users to specify some of these attributes of a data item, e.g. OpenMP/OpenACC {\it shared} and {\it private} clause for specifying the 1) shared-private attribute, OpenMP {\it map} clause and OpenACC {\it copyin/copy/copyout} clause for the 2) mapping attribute, and the {\it allocate} and {\it alloc} clause in OpenMP and related clause in OpenACC for specifying the 5) allocator attribute. For those attributes that are not explicitly specified in a program, the language applies default rules that the compiler can use to append the corresponding attributes. 


\REM{

The mm attribute is used for the user to specify a memory allocator and de-allocator for a data item when it is mapped or privatized if users want the memory of the data item on the target to be in a specified location such as HBM, PMEM or DRAM, or to use a customized memory allocator. 
For the attribute for data access modes, which is used to indicate whether a data item is accessed as \textit{read-only}, \textit{write-only}, or \textit{read-write}, it will reply on compiler analysis to collect that information since most programming models do not provide constructs for the user to annotate that information. 



The data distribution attribute is included to support partitioning the array into array sections among computing units. Having this attribute is motivated by parallel programming models that aim for high programming productivity for HPC, such as the early High Performance Fortran (HPF) effort, and PGAS efforts such as UPC, Chapel and Co-Array Fortran that support data distribution. Recent OpenMP standard allows using array sections in specifying data mapping items, and related work of exploring data distributions and computation-data binding on OpenMP~\cite{yan2017homp} are all relevant work that demonstrate the advantage of data distribution. 
}

\subsection{UPIR for Explicit Data Movement and Memory Management}
\label{subsec:pfgspec_data_movement}
The IRs for data movement operations are used for specifying the actual operations that would incur moving data from one location to another. Such operation could be explicitly specified in a program using language-provided constructs, such as the {\it update} directive in OpenACC and {\it target update} directive in OpenMP, or analyzable by compilers for known data-movement APIs such as {\it memcpy} or {\it cudaMemCpy}. 


Similar to the data movement operations, IRs for memory management operations specify the memory allocator and de-allocator supported by language constructs or those operations that can be analyzed by compilers for API calls such as {\it malloc}, {\it cudaMalloc}, {\it hbm\_alloc}, {\it pmem\_alloc}, etc.  
Making those operations as part of the IR and to be analyzable would facilitate the optimization and code transformation. 
Users can specify the information to guide the compiler to allocate the data to desired locations.
For example, a huge array can be allocated to the memory space with high capacity by using \textit{allocator (large\_cap\_mem\_alloc)} so that it will not cause the out-of-memory error.
The UPIR specification for both data movement and mm are provided in Figure~\ref{fig:pfg_mlir_data_movement}. 


\begin{figure}[h!]
\begin{lstlisting}[frame=single]
upir.data_movement ::= 'data_movement' '(' dest-target, dest-ptr, src-target src-ptr, dm-size ')'  dm-direction data-memcpy dm-field-list
dest-target ::= expr_id
...
dm-direction ::== forward|backward //to allow two direction of data movement
dm-field-list ::= dm-field | dm-field dm-field-list
dm-field ::= depend | ...

//upir.data_update is simplified data movement IR. 
upir.data_update ::= 'data_update' '(' data-list ')' dm-direction data-memcpy dm-field-list
data-list ::= expr_id-list
upir.mm-allocator ::= 'mm_allocator' '(' allocator-attr ')'
upir.mm-deallocator ::= 'mm_deallocator' '(' deallocator-attr ')'
\end{lstlisting}
\caption{UPIR MLIR dialect for explicit data movement and memory management}
\label{fig:pfg_mlir_data_movement}
\vspace{-0.3cm}
\end{figure}


In summary, with this UPIR design for data attributes, data movement and memory management operations, they are able to provide adequate information for the compiler to enable advanced data-flow and data-aware analysis and optimization for parallel programs. There are two guidelines of the design to aid that compiler transformation and optimization: 1) data attributes provide a rich set of information about data sharing, but leaving data movement and memory management operations as compiler optimizations for the purpose of achieving maximum overlapping and pipelining of computation and data movement, the fusion of data movement and memory allocation, etc. 2) data movement and memory management operations explicitly specified in the program are also analyzable and optimizable with regards to other operations including computation and implicit operations of them rendered by the data attributes of data items, and synchronizations such as barrier and reductions.

%% file: src/upir-sync.tex
This group of IR elements is for representing the language constructs and APIs that prescribe the behaviors of communications and coordination operations between parallel work units. We categorize those operations into three sub-groups: 1) those that involve all participating units, referred to as collectives such as barrier (OpenMP barrier, MPI\_Barrier, etc), broadcast (MPI\_Bcast), reduction in OpenMP/OpenACC/MPI, etc; and 2) those that involve only two units, namely point-to-point (p2p) operations, such as data shuffling between threads and message passing between two MPI processes, and 3) mutual exclusion and locks/unlocks that involves collective participation but one primary unit at a time such as OpenMP single, atomic, etc. For most of those operations, there could be synchronous and asynchronous versions. Using asynchronous operations helps achieve overlapping of synchronization/communication with computations. 

Many of those operations are provided or implemented as runtime APIs, thus compiler transformation of those operations could be simply converting the languages constructs to runtime API. For optimization, previous work has shown that the compiler can optimize the use of this group of constructs, such as reducing redundant barriers or global synchronizations~\cite{chen1999redundant,midkiff1987compiler}, compiler-assisted optimization of MPI calls~\cite{danalis2009mpi, friedley2011communication}, and optimizing mutual excluded code sections that use heavy locks~\cite{novillo2000optimizing}. 
Converting synchronous operations to asynchronous ones by the compiler is also an effective way of optimization for the synchronizations for parallel programs~\cite{prakash1993synchronization, shirako2009chunking}. 
Thus adequately representing those constructs in compiler IR is necessary for compiler transformation and parallelism-aware optimizations.

The design of UPIR elements for synchronization operations needs to consider two aspects: 1) to unify the IRs for various types of synchronizations, and 2) to unify the IRs for synchronous and asynchronous synchronizations. 
For the first aspect, we consider four fields that are common among the various types of sync operations: 1) the primary unit that participates in the operation, e.g. the thread that collects the results in reduction operation, or the source process in an MPI\_Bcast operation, 
2) the secondary unit (s) that participate in the operation, e.g. the receivers of a broadcast or an MPI\_Recv, 3) the computation or operation that is performed with the synchronization, e.g. reduction or broadcasting, and 4) the data that is used in the synchronization. 

For the second aspect, which is to unify the IRs for the synchronous and asynchronous versions of each operation,
we consider two steps when a sync operation, particularly collective syncs, are performed. The first step is arrive-compute, which indicates that a work unit arrives at the sync point and performs the necessary computation or operations such as an addition in an add-reduction, or sending or receiving messages for broadcast operations. The second step is wait-release, which indicates that a work unit waits for the synchronization to be performed by all participating units and then is released to continue. For synchronous synchronization, the two steps are performed by one API call. For asynchronous version, two API calls, one for each step are performed, thus allowing adding computation between these two calls to achieve overlapping of synchronization with computation. These two steps are also similar to lock and unlock operations for mutual exclusion operations. 
With these two aspects of unification, we believe the designed IR would facilitate much more uniform compiler passes for the optimization of synchronizations than using independent IRs for each operation. 
Following this design, we describe the specification of the IRs for the synchronizations in Figure~\ref{fig:pfg_mlir_synchronization}.


\begin{figure}[h!]
\begin{lstlisting}[frame=single]
upir.sync ::= sync-name sync-async primary secondary operation data-list implicit
sync-name ::= 'barrier' | 'reduction' | 'taskwait' | 'broadcast' | 'allreduce' | 'send' | 'recv' | 'single' | critical | atomic
sync-async ::= 'sync' | 'async' step
step ::= 'arrive-compute' | 'wait-release'
primary ::= 'primary' '(' sync-unit ')'
secondary ::= 'primary' '(' sync-unit ')'
sync-unit ::= 'task' | 'thread' | 'rank' ':' unit_id
unit-id ::= expr_id | '*'
operation ::= //sync-specific operation, e.g. add for add-reduction
data-list ::= expr-id-list
implicit ::= | 'implicit'
\end{lstlisting}
\caption{UPIR MLIR dialect for synchronization}
\label{fig:pfg_mlir_synchronization}
\vspace{-0.2cm}
\end{figure}

\REM{

\subsubsection{Synchronous Collectives: Barrier and Reduction}
\label{subsec:synchronization}
Barrier and reduction are two commonly used synchronous collectives in lots of programming languages including OpenMP, OpenACC, MPI, CUDA, etc. Barrier can be considered as simplified synchronized collectives that also include operations such as broadcast (MPI\_Bcast), etc. 
PFG always creates synchronization IR explicitly, but with an attribute to tell whether it is implicit from the user's point of view (Fig.~\ref{fig:pfg_mlir_synchronization}).
For example, given a parallel for loop without \texttt{nowait} clause, there is an implicit barrier right after the loop even the user doesn't specify this information.
Besides the worksharing IR, PFG also appends a barrier IR to the loop.
If another synchronization is explicitly declared here, there would be two barriers and the compiler can easily recognize that one of them can be eliminated to prevent redundancy.

\begin{figure}[ht!]
\begin{lstlisting}[frame=single]
upir.barrier (::mlir::barrier::BarrierOp)
operation ::= 'barrier' clause-list
clause-list ::= clause | clause clause-list
clause ::= task | implicit | cost
task ::= 'task' '(' task-id ')'
task-id ::= ssa-id
implicit ::= 'implicit'
\end{lstlisting}
\caption{UPIR dialect in MLIR for synchronization}
\label{fig:pfg_mlir_synchronization}
\end{figure}

\subsubsection{Reduction}


\begin{figure}[ht!]
\begin{lstlisting}[frame=single]
reduction ::= 'reduction' '(' reduction-parameter ',' reduction-mode ':' data-list ')'
reduction-mode ::= 'all-task' | 'in-task' | 'all-unit'
reduction-parameter ::= modifier ',' identifier  | identifier
modifier ::= 'inscan' | 'task' | 'default'
identifier ::= '+' | '-' | '*' | ...
\end{lstlisting}
\caption{UPIR dialect in MLIR for reduction}
\label{fig:pfg_mlir_reduction}
\end{figure}

Reduction clause is used to reduce partial results from certain threads into one by specified operations.
PFG unifies the reduction operation in different scenarios into one IR (Fig.~\ref{fig:pfg_mlir_reduction}).
Besides the conventional reduction operators, such as addition and multiplication, a field named \texttt{reduction-mode} is provided to indicate the threads participating in the reduction.
\texttt{all-task} and \texttt{in-task} present all the tasks or selected tasks join the reduction.
\texttt{all-unit}, as the default value of this field, will reduce the specified variables from all SPMD units.

\subsubsection{Mutual exclusion}

Instead of creating individual IRs, PFG proposes a high-level abstraction for mutual exclusion (Fig.~\ref{fig:pfg_mlir_mutual_exclusion}).
Any parallel task or SPMD branch can specify the information of mutual exclusion and its followed code body will act accordingly.
For instance, \texttt{task mutual-exclusion(single)} indicates that the adjacent code block must be executed on exactly only one SPMD unit once.

\subsubsection{Taskwait}. this could be in 3.3

\begin{figure}[ht!]
\begin{lstlisting}[frame=single]
mutual-exclusion ::= 'mutual-exclusion' '(' mutual-exclusion-parameter ')'
mutual-exclusion-parameter ::= 'critical' | 'atomic' | 'single' | 'master'
\end{lstlisting}
\caption{UPIR dialect in MLIR for mutual exclusion}
\label{fig:pfg_mlir_mutual_exclusion}
\end{figure}
}





%% file: src/eva.tex
In this section, we present a prototype implementation of the UPIR in the design and describe how the UPIR facilitates compiler transformation. 
We show how UPIR supports the source code in CUDA, C/C++, and Fortran with OpenMP/OpenACC. The performance evaluation is conducted by comparing our compiler with the OpenMP and OpenACC compilers of LLVM, GCC and NVIDIA.



\begin{figure*}[!htb]
\centering
\includegraphics[width=0.84\linewidth]{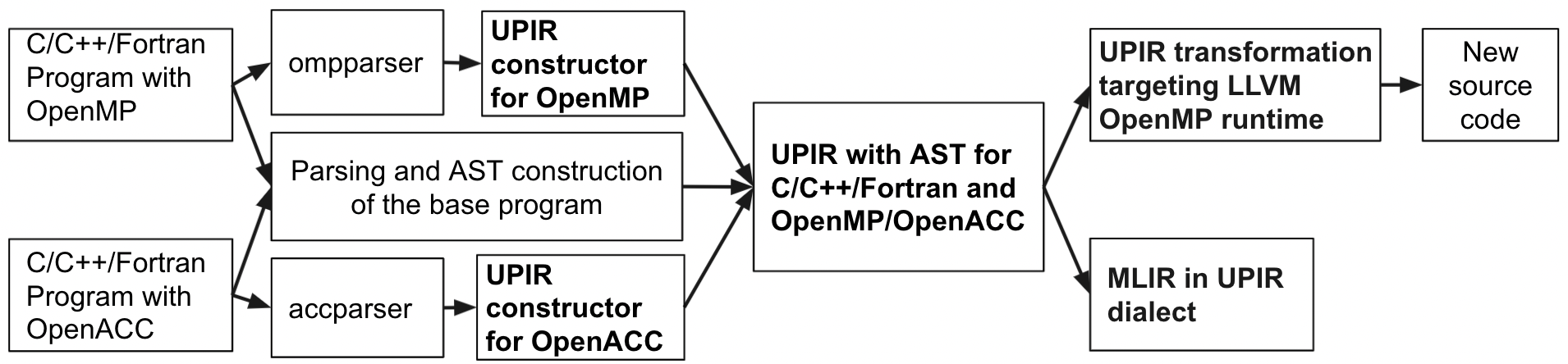}
\caption{UPIR implementation in ROSE compiler to support C/C++/Fortran and OpenMP and OpenACC}
\label{upir_arch}
\end{figure*} 


Our prototype is implemented in ROSE source-to-source compiler.
ROSE compiler supports C/C++, Fortran, Java, CUDA, and several other languages.
It provides source-to-source transformation and a rich set of APIs for program transformation for quick prototyping with high quality. It enables us to implement the UPIR and the support for multiple parallel models much more productively than any other compiler framework such as GCC or LLVM.
OpenACC is a parallel programming language that is similar to OpenMP but more focused on accelerators.
The original ROSE compiler does not support OpenACC. However, by adding an independent OpenACC directive parser to ROSE, we can handle OpenACC source code and convert it to UPIR. 
The most significant advantage of using UPIR for both is that they can share a unified transformation.

Figure~\ref{upir_arch} shows how UPIR is generated from OpenMP and OpenACC source code, in both C and Fortran, and followed by a unified transformation.
ROSE uses EDG and Open Fortran parser to parse the C/C++ and Fortran source code, respectively.
We integrate a separate OpenMP parser ompparser~\cite{wang2019ompparser} and an OpenACC accparser into ROSE to parse OpenMP and OpenACC directives.
Their parser IRs are converted to the unified UPIR, regardless of the base language of the source code.
A data analysis module is implemented to collect explicit and implicit data usage information, and populates the UPIRs with the complete data attribute.  

The UPIR is also implemented with LLVM TableGen to produce the UPIR dialects in MLIR, allowing the ROSE implementation of the UPIR to be exported to MLIR (Figure \ref{fig:axpy_omp_acc_mlir_output}, \ref{fig:axpy_cuda_mlir_output}).
Thus, other than the transformation for UPIR in the ROSE compiler, developers can also work on the exported MLIR using the LLVM toolchain.
The exported MLIR includes the same parallelism information as our UPIR implementation in ROSE, e.g. data attributes and movements in a parallel region (lines 3-6 in Figure~\ref{fig:axpy_omp_acc_mlir_output}).
However, LLVM and MLIR are more complicated than ROSE and require much more effort to implement the transformation of the supported feature. 

\begin{figure}[!htb]
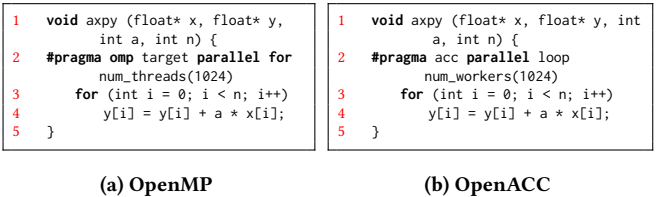

\begin{lrbox}{\codebox}%
\begin{minipage}[b]{0.47\linewidth}%
\begin{lstlisting}[frame=single]
void axpy (float* x, float* y, int a, int n) {
#pragma omp target parallel for num_threads(1024)
    for (int i = 0; i < n; i++)
        y[i] = y[i] + a * x[i];
}
\end{lstlisting}%
\end{minipage}%
\end{lrbox}%
\subfloat[OpenMP]{\usebox{\codebox}}%
\hfill
\begin{lrbox}{\codebox}%
\begin{minipage}[b]{0.49\linewidth}%
\begin{lstlisting}[frame=single]
void axpy (float* x, float* y, int a, int n) {
#pragma acc parallel loop num_workers(1024)
    for (int i = 0; i < n; i++)
        y[i] = y[i] + a * x[i];
}
\end{lstlisting}
\end{minipage}%
\end{lrbox}%
\subfloat[OpenACC]{\usebox{\codebox}}%
\caption{AXPY in OpenMP and OpenACC for GPU offloading}
\label{fig:axpy_src_opemp_openacc}
\end{figure}

\begin{figure}[htb!]
\begin{lstlisting}[frame=single]
func @axpy(%arg0: memref<*xi32, 8>, %arg1: memref<*xi32, 8>, %arg2: i32, %arg3: i32) {
  ... // %2, %3, %4, %5 are the data used in the parallel region
  %2 = upir.parallel_data_info(x, shared, implicit, tofrom, implicit, read-only)
  %3 = upir.parallel_data_info(y, shared, implicit, tofrom, implicit, read-write)
  %4 = upir.parallel_data_info(a, shared, implicit, tofrom, implicit, read-only)
  %5 = ...
  %c6_i32 = constant 1024 : i32
  upir.task target(nvptx) data(%2, %3, %4, %5) {
    upir.spmd num_units(%c6_i32 : i32) data(%2, %3, %4, %5) target(gpu) {
      %c0 = constant 0 : index
      %c1 = constant 1 : index
      upir.loop induction-var(%arg4) lowerBound(%c0) upperBound(%arg3) step(%c1) {
        upir.loop-parallel worksharing {
          ...
} } } } }
\end{lstlisting}
\caption{AXPY in UPIR MLIR dialect, for OpenMP and OpenACC GPU Offloading of Figure~\ref{fig:axpy_src_opemp_openacc}}
\label{fig:axpy_omp_acc_mlir_output}
\end{figure}

\begin{figure}[hb!]
\begin{lstlisting}[frame=single]
func @axpy(%arg0: memref<f64>, %arg1: memref<f64>, %arg2: f64, %arg3: i32) {
  %c6_i32 = constant 1024 : i32
  acc.parallel num_workers(%c6_i32: i32) {
    %c0 = constant 0 : index
    %c1 = constant 1 : index
    acc.loop worker {
      scf.for %arg4 = %c0 to %arg3 step %c1 {
        ...
} } } }
\end{lstlisting}
\caption{AXPY in OpenACC MLIR dialect translated from UPIR shown in Figure~\ref{fig:axpy_omp_acc_mlir_output}}
\label{fig:axpy_pir_acc_mlir_output}
\end{figure}

Given the OpenMP and OpenACC versions of AXPY in Figure~\ref{fig:axpy_src_opemp_openacc}, identical UPIRs are generated (Figure~\ref{fig:axpy_omp_acc_mlir_output}) because they both reveal the same parallelism information.
Our implementation produces the same lowered source code from these two inputs.
It saves much effort by not developing different lowering modules for the two languages. 
If necessary, language-specific transformations can be appended after the common ones.
The exported MLIR in UPIR dialect can be translated to other dialects or cooperate with them.
For instance, the UPIR of AXPY above can be converted to MLIR OpenACC dialect (Figure~\ref{fig:axpy_pir_acc_mlir_output}).
Developers who are more familiar with OpenACC can apply their optimizations upon OpenACC MLIR translated from UPIR.

\subsection{Using UPIR to Represent CUDA Kernel and Launching}


CUDA is another popular parallel programming language designed for NVIDIA GPUs.
However, unlike OpenMP and OpenACC, which use simpler directive annotations, it requires additional effort to learn the programming APIs.
The language-independent design of UPIR can take both their advantages.
UPIR can represent the parallelism and data usage of CUDA kernel calls.
Figure~\ref{fig:axpy_src_cuda} and \ref{fig:axpy_cuda_mlir_output} show a CUDA version of AXPY and its UPIR.
The \texttt{task} IR with \texttt{device} attribute indicates that the kernel runs on NVIDIA GPU.
\texttt{num\_teams} and \texttt{num\_units} attributes of inner \texttt{spmd} IR are corresponding to blocks and threads of the CUDA kernel.
The \texttt{task} and \texttt{spmd} IRs are always perfectly nested since they are converted from one CUDA kernel call.

\begin{figure}[hb!]
\begin{lstlisting}[frame=single]
__global__ void axpy_kernel(float* x, float* y, int a, int n) {
    int i = blockDim.x * blockIdx.x + threadIdx.x;
    if (i < n) y[i] = y[i] + a * x[i];
}
void axpy(float* d_x, float* d_y, int a, int n) {
    axpy_kernel<<<(n+255)/256, 256>>>(d_x, d_y, a, n);
}
\end{lstlisting}
\caption{AXPY source code in CUDA}
\label{fig:axpy_src_cuda}
\end{figure}

\begin{figure}[hb!]
\begin{lstlisting}[frame=single]
func @axpy_kernel(%arg0: memref<*xi32, 8>, %arg1: memref<*xi32, 8>, %arg2: i32, %arg3: i32) { ... }
func @axpy(%arg0: memref<*xi32, 8>, %arg1: memref<*xi32, 8>, %arg2: i32, %arg3: i32) {
  ... // %2, %3, %4, %5 are the data used in the parallel region
  upir.task device(nvptx) data(%2, %3, %4, %5) {
    upir.spmd num_teams(%1 : i32) num_units(%c256_i32_0 : i32) data(%2, %3, %4, %5) target(gpu) {
      call @axpy_kernel(%arg0, %arg1, %arg2, %arg3) : (memref<*xi32, 8>, memref<*xi32, 8>, i32, i32) -> ()
} } }
\end{lstlisting}
\caption{AXPY in UPIR MLIR dialect, for CUDA of Figure~\ref{fig:axpy_src_cuda}}
\label{fig:axpy_cuda_mlir_output}
\end{figure}

The produced UPIR can be unparsed to other parallel programming languages, such as OpenMP.
By doing so, we can run CUDA kernels on CPU and other computing devices. In addition, the new source code in OpenMP or OpenACC is also easier to program.
Base on the same concept, it is also possible to lower certain UPIRs to CUDA source code.

\subsection{Performance Evaluation of using UPIR in Compiler Transformation for OpenMP and OpenACC Offloading}

UPIR aims to unify the representation of parallelism across multiple parallel programming languages. 
It helps compilers conduct a unified transformation for multiple parallel programming models.
Therefore, our performance demonstration shows that implementing transformations on top of UPIR in a compiler can support both OpenMP and OpenACC.
We pick four offloading kernels for evaluation: AXPY, matrix multiplication, matrix-vector multiplication, and 2D stencil.
They are implemented in both OpenMP and OpenACC. 
The uses of OpenMP and OpenACC directives in these kernels represent the commonly used parallel constructs of these two models in many applications~\cite{loff2021parallel}. These examples demonstrate the use of UPIR for SPMD (OpenMP's \textit{parallel} and \textit{teams}, OpenACC's \textit{parallel}), data parallelism (OpenMP's \textit{distribute} and \textit{for}, and OpenACC's \textit{loop}), offloading tasks (OpenMP's \textit{target}, OpenACC's \textit{parallel}), data attributes (OpenMP's \textit{map}, {\it private}, {\it shared}, \textit{target data} and OpenACC's \textit{data} and \textit{copyin/copyout/copy}), sync (OpenMP's \textit{barrier}, OpenACC's \textit{wait}), etc. For each kernel, we experimented a large range of problem sizes for performance collection, but only reported in the paper the problem sizes that sufficiently demonstrate the performance trends. Each kernel is executed 10 times and the collected execution time is the average of the 10 execution.   
Thus, we believe selected kernels are representative enough to serve the purpose.
The evaluation compared our ROSE-based UPIR compiler, NVIDIA HPC SDK, and GCC compile, all for both OpenMP and OpenACC, and 
Clang/LLVM for OpenMP only since LLVM does not support OpenACC.
The execution time is presented in log-scale for better readability.
Our experimental platform has 2 CPUs (20 cores for each), 512 GB of RAM, and one NVIDIA V100 GPU with 32 GB of HBM.
The system runs Clang/LLVM 14.0, NVIDIA HPC SDK 22.1 with CUDA toolkit 11.5, and GCC 11.2 on Ubuntu 20.04. All compilations enable -O3 flag.

\REM {
The hardware and software configurations for the evaluation are listed in Tab.~\ref{tab:machines}.

\begin{table}[!ht]
	\centering
\begin{tabular}{|l|l|}  \hline
CPU        &  Intel Xeon Gold 6230N CPU 2.30GHz \\ \hline
Cores      & 2 sockets $\times$ 20 physical cores  \\ \hline 
Main Mem   & 256 GB    \\ \hline
GPU        & Nvidia Tesla V100   \\ \hline
Device Mem & 32 GB           \\ \hline
OS         & Ubuntu 18.04 LTS  \\ \hline
Compilers  & Clang/LLVM 12.0.1, PGI 20.11 \\ \hline 
CUDA       & 11.2               \\ \hline
\end{tabular}
\caption{Experimental platform}
\label{tab:machines}
\end{table}
}

\begin{figure}[!htb]
\centering
\resizebox{\linewidth}{!}{%
\input{figs/rex-performance-axpy-carina}
}
\caption{AXPY performance of UPIR compiler, LLVM, NVIDIA, and GCC compilers}
\label{upir_axpy_performance_carina}
\end{figure}
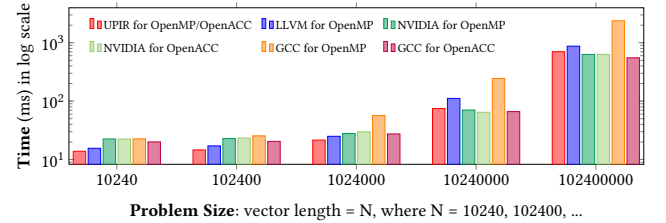

\begin{figure}[!htb]
\centering
\resizebox{\linewidth}{!}{%
\input{figs/rex-performance-matmul-carina}
}
\caption{Matrix multiplication performance of UPIR compiler, LLVM, NVIDIA, and GCC compilers}
\label{upir_matmul_performance_carina}
\end{figure}

\begin{figure}[!htb]
\centering
\resizebox{\linewidth}{!}{%
\input{figs/rex-performance-matvec-carina}
}
\caption{Matrix-vector multiplication performance of UPIR compiler, LLVM, NVIDIA, and GCC compilers}
\label{upir_matvec_performance_carina}
\end{figure}

\begin{figure}[!htb]
\centering
\resizebox{\linewidth}{!}{%
\input{figs/rex-performance-stencil-carina}
}
\caption{2D stencil performance of UPIR compiler, LLVM, NVIDIA, and GCC compilers}
\label{upir_stenicl_performance_carina}
\end{figure}
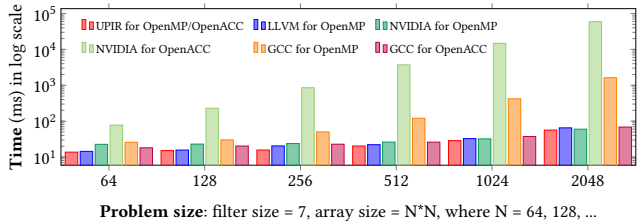

The performance results of the four compilers (UPIR, NVIDIA, GCC for OpenMP/OpenACC, and LLVM for OpenMP) are shown in Figure~\ref{upir_axpy_performance_carina}, \ref{upir_matmul_performance_carina}, \ref{upir_matvec_performance_carina}, \ref{upir_stenicl_performance_carina}.
For the OpenMP version, our implementation can achieve up to 1.28x speedup over LLVM and 25.89x speed up over GCC in average for all the problem sizes we selected.
For the OpenACC version, UPIR shows up to 235.1x speedup over NVIDIA compiler and 1.15x speedup over GCC in average for all the evaluated problem sizes.
We believe LLVM considers more general cases of offloading, thus might introduce more overhead in kernel launching and thread management internally. For example, LLVM use the state machine~\cite{doerfert2019tregion,bertolli2014coordinating}
to manage dynamic threading on NVIDIA GPU with relatively heavy overhead.
Thus, we use dynamic parallelism introduced in CUDA 5.0 for threading management.
It directly launches the number of threads for nested kernels as demanded on the device.
It prevents the communication overhead to the host and does not need to maintain the state of redundant threads.
This demonstration shows that UPIR can assist in the unified compiler transformation.
It significantly reduces the compiler programming effort to support multiple parallel programming models because we do not need to maintain several similar versions of compiler transformation.
Moreover, UPIR does not sacrifice performance to achieve this goal.


\subsubsection{Performance Analysis}
We conducted detailed performance analysis of the compilers 
for OpenMP and OpenACC programs. GCC utilizes the unified IR GIMPLE and maps both OpenMP and OpenACC code to GIMPLE for later transformation.
Theoretically, if they are semantic equivalent, the compiled executable should lead to the same performance.
However, in all four kernels, the OpenACC version outperforms the OpenMP version significantly.
We notice that all the number of GPU threads per block used in the kernels are limited to 256 even 1024 is specified in the source code.
This issue only exists in the OpenMP version but not OpenACC version.
GCC always follows the specified number of threads in the OpenACC code.
Although GCC adopts the idea of unified parallel IR, it fails to deliver a consistent performance for the same kernel in different languages.

The similar problem happens to LLVM.
LLVM determines the maximum number of threads supported by GPU.
If the specified value exceeds the limit, it will be reduced to that threshold.
However, in our experiments, LLVM does not always obtain the threshold correctly.
It may use different number of threads instead.

The latest NVIDIA HPC SDK supports both OpenMP and OpenACC programs.
It uses different sets of IR for compilation, such as \texttt{\_\_nvomp\_*} in OpenMP and \texttt{\_\_pgi\_uacc\_*} in OpenACC.
Our implementation of OpenMP and OpenACC code are semantic equivalent.
Thus, it is supposed to lead to a quite similar executable generated from the same compiler.
However, NVIDIA compiler does not always show a consistent performance.
For stencil and matrix multiplication, the OpenACC version is much slower than the OpenMP version and also slower than the executable built by other compilers.
The profiling shows that the generated GPU code of stencil in OpenACC takes about 71 millions cycles while the code generated by LLVM only takes around 50 thousands cycles.
About 99\% of elapsed kernel time is spent on \texttt{\_\_acc\_wait} for synchronization.
Additionally, NVIDIA compiler is able to compile the OpenMP version of these two test cases, but the execution results in either a kernel launching failure or incorrect computation.
It indicates that the transformation for OpenMP and OpenACC in NVIDIA compiler is not unified.

To sum up, GCC and NVIDIA compilers support both OpenMP and OpenACC, and we notice some unification in their IR and their transformations for different parallel programming models.
However, they do not deliver a consistent performance for the OpenMP and OpenACC programs of the same computation and parallel semantics.
In contrast, UPIR compiler is able to support OpenMP/OpenACC code with unified transformation and achieves the same performance.

%% file: figs/rex-performance-axpy-carina.tex
\begin{tikzpicture}
 
\begin{axis} [
    ybar,
    xlabel={\textbf{Problem Size}: vector length = N, where N = 10240, 102400, ...},
    ylabel={\textbf{Time} (ms) in log scale},
    width=\textwidth,
    height=6cm,
    legend style={nodes={scale=1, transform shape},
    at={(0.4,0.95)}, anchor=north,
    legend columns=-1,
    /tikz/every even column/.append style={column sep=6pt},
    fill opacity=0, text opacity=1, row sep=6pt, draw=none},
    symbolic x coords={10240,102400,1024000,10240000,102400000},
    ymode=log,
    log base y ={10},
    legend cell align=left,
    legend columns = 3,
    xtick distance = 1,
    label style={font=\LARGE},
    tick label style={font=\LARGE},
    x label style={at={(axis description cs:0.5,-0.1)},anchor=north},
    ]

      \addplot[red, fill=red, fill opacity=0.5] plot coordinates {
      (10240, 13.85) 
      (102400, 14.61) 
      (1024000, 21.54) 
      (10240000, 74.75) 
      (102400000, 702.39) 
      };
      \addplot[blue, fill=blue, fill opacity=0.5] plot coordinates {
      (10240, 15.58) 
      (102400, 17.12) 
      (1024000, 25) 
      (10240000, 111.36) 
      (102400000, 873.46) 
      };
      \addplot[ForestGreen, fill=ForestGreen, fill opacity=0.5] plot coordinates {
      (10240, 22.5) 
      (102400, 22.87) 
      (1024000, 27.86) 
      (10240000, 70.4) 
      (102400000, 630.29) 
      };
      \addplot[YellowGreen, fill=YellowGreen, fill opacity=0.5] plot coordinates {
      (10240, 22.43) 
      (102400, 23.3) 
      (1024000, 29.78) 
      (10240000, 64.47) 
      (102400000, 631.13) 
      };
      \addplot[orange, fill=orange, fill opacity=0.5] plot coordinates {
      (10240, 22.62) 
      (102400, 25.46) 
      (1024000, 56.62) 
      (10240000, 244.89) 
      (102400000, 2373.74) 
      };
      \addplot[purple, fill=purple, fill opacity=0.5] plot coordinates {
      (10240, 20.02) 
      (102400, 20.45) 
      (1024000, 27.44) 
      (10240000, 66.18) 
      (102400000, 552.74) 
      };
      \legend{UPIR for OpenMP/OpenACC, LLVM for OpenMP, NVIDIA for OpenMP, NVIDIA for OpenACC, GCC for OpenMP, GCC for OpenACC}

\end{axis}
\end{tikzpicture}

%% file: figs/rex-performance-matmul-carina.tex
\begin{tikzpicture}
 
\begin{axis} [
    ybar,
    xlabel={\textbf{Problem Size}: matrix size = N*N, where N = 64, 128, ...},
    ylabel={\textbf{Time} (ms) in log scale},
    width=\textwidth,
    height=6cm,
    legend style={nodes={scale=1, transform shape},
    at={(0.4,0.95)}, anchor=north,
    legend columns=-1,
    /tikz/every even column/.append style={column sep=6pt},
    fill opacity=0.7, text opacity=1, row sep=6pt, draw=none},
    symbolic x coords={64,128,256,512,1024},
    ymode=log,
    log base y ={10},
    legend cell align=left,
    legend columns = 3,
    xtick distance = 1,
    label style={font=\LARGE},
    tick label style={font=\LARGE},
    x label style={at={(axis description cs:0.5,-0.1)},anchor=north},
    ]

      \addplot[red, fill=red, fill opacity=0.5] plot coordinates {
      (64, 13.91) 
      (128, 17.49) 
      (256, 27.6) 
      (512, 135.03) 
      (1024, 976.88) 
      };
      \addplot[blue, fill=blue, fill opacity=0.5] plot coordinates {
      (64, 13.81) 
      (128, 16.25) 
      (256, 28.73) 
      (512, 136) 
      (1024, 1011.05) 
      };
      \addplot[YellowGreen, fill=YellowGreen, fill opacity=0.5] plot coordinates {
      (64, 34.75) 
      (128, 75.41) 
      (256, 238.99) 
      (512, 944.62) 
      (1024, 4423.29) 
      };
      \addplot[orange, fill=orange, fill opacity=0.5] plot coordinates {
      (64, 22.54) 
      (128, 34.78) 
      (256, 116.63) 
      (512, 802.32) 
      (1024, 5847.54) 
      };
      \addplot[purple, fill=purple, fill opacity=0.5] plot coordinates {
      (64, 22.4) 
      (128, 22.78) 
      (256, 36.83) 
      (512, 147.28) 
      (1024, 1012.82) 
      };
      \legend{UPIR for OpenMP/OpenACC, LLVM for OpenMP, NVIDIA for OpenACC, GCC for OpenMP, GCC for OpenACC}

      \end{axis}
    \end{tikzpicture}

%% file: figs/rex-performance-matvec-carina.tex
    \begin{tikzpicture}
      \begin{axis}[
    ybar,
    xlabel={\textbf{Problem Size}: vector length = N, matrix size = N*N, where N = 1024, 2048, ...},
    ylabel={\textbf{Time} (ms) in log scale},
    width=\textwidth,
    height=6cm,
    legend style={nodes={scale=1, transform shape},
    at={(0.4,0.95)}, anchor=north,
    legend columns=-1,
    /tikz/every even column/.append style={column sep=6pt},
    fill opacity=0, text opacity=1, row sep=6pt, draw=none},
    symbolic x coords={1024, 2048, 4096, 8192, 16384},
    ymode=log,
    log base y ={10},
    legend cell align=left,
    legend columns = 3,
    xtick distance = 1,
    label style={font=\LARGE},
    tick label style={font=\LARGE},
    x label style={at={(axis description cs:0.5,-0.1)},anchor=north},
    ]

      \addplot[red, fill=red, fill opacity=0.5] plot coordinates {
      (1024, 17.88) 
      (2048, 20.7) 
      (4096, 44.2) 
      (8192, 145.74) 
      (16384, 583.45) 
      };
      \addplot[blue, fill=blue, fill opacity=0.5] plot coordinates {
      (1024, 24.32) 
      (2048, 30.08) 
      (4096, 43.81) 
      (8192, 146.51) 
      (16384, 582.3) 
      };
      \addplot[ForestGreen, fill=ForestGreen, fill opacity=0.5] plot coordinates {
      (1024, 22.77) 
      (2048, 24.58) 
      (4096, 30.49) 
      (8192, 66.83) 
      (16384, 218.13) 
      };
      \addplot[YellowGreen, fill=YellowGreen, fill opacity=0.5] plot coordinates {
      (1024, 26.09) 
      (2048, 31.56) 
      (4096, 44.68) 
      (8192, 100.95) 
      (16384, 282.31) 
      };
      \addplot[orange, fill=orange, fill opacity=0.5] plot coordinates {
      (1024, 27.98) 
      (2048, 54.53) 
      (4096, 161.14) 
      (8192, 607.94) 
      (16384, 2399.33) 
      };
      \addplot[purple, fill=purple, fill opacity=0.5] plot coordinates {
      (1024, 21.53) 
      (2048, 26.17) 
      (4096, 47.1) 
      (8192, 145.47) 
      (16384, 539.2) 
      };
      \legend{UPIR for OpenMP/OpenACC, LLVM for OpenMP, NVIDIA for OpenMP, NVIDIA for OpenACC, GCC for OpenMP, GCC for OpenACC}

      \end{axis}
    \end{tikzpicture}

%% file: figs/rex-performance-stencil-carina.tex
\begin{tikzpicture}
 
\begin{axis} [
    ybar,
    xlabel={\textbf{Problem size}: filter size = 7, array size = N*N, where N = 64, 128, ...},
    ylabel={\textbf{Time} (ms) in log scale},
    width=\textwidth,
    height=6cm,
    legend style={nodes={scale=1, transform shape},
    at={(0.4,0.95)}, anchor=north,
    legend columns=-1,
    /tikz/every even column/.append style={column sep=6pt},
    fill opacity=0, text opacity=1, row sep=6pt, draw=none},
    symbolic x coords={64,128,256,512,1024,2048},
    ymode=log,
    log base y ={10},
    legend cell align=left,
    legend columns = 3,
    xtick distance = 1,
    label style={font=\LARGE},
    tick label style={font=\LARGE},
    x label style={at={(axis description cs:0.5,-0.1)},anchor=north},
    ]

      \addplot[red, fill=red, fill opacity=0.5] plot coordinates {
      (64, 13.69) 
      (128, 15.14) 
      (256, 15.7) 
      (512, 20.32) 
      (1024, 28.72) 
      (2048, 56.47) 
      };
      \addplot[blue, fill=blue, fill opacity=0.5] plot coordinates {
      (64, 14.39) 
      (128, 15.68) 
      (256, 20.43) 
      (512, 22.13) 
      (1024, 32.65) 
      (2048, 65.08) 
      };
      \addplot[ForestGreen, fill=ForestGreen, fill opacity=0.5] plot coordinates {
      (64, 22.63) 
      (128, 22.95) 
      (256, 23.75) 
      (512, 26.09) 
      (1024, 32.07) 
      (2048, 59.82) 
      };
      \addplot[YellowGreen, fill=YellowGreen, fill opacity=0.5] plot coordinates {
      (64, 77.77) 
      (128, 228.96) 
      (256, 853.47) 
      (512, 3709.49) 
      (1024, 14805) 
      (2048, 59176) 
      };
      \addplot[orange, fill=orange, fill opacity=0.5] plot coordinates {
      (64, 25.73) 
      (128, 30.06) 
      (256, 50.18) 
      (512, 120.3) 
      (1024, 420.18) 
      (2048, 1624.27) 
      };
      \addplot[purple, fill=purple, fill opacity=0.5] plot coordinates {
      (64, 18.07) 
      (128, 20.24) 
      (256, 22.79) 
      (512, 26.05) 
      (1024, 37.42) 
      (2048, 68.48) 
      };
      \legend{UPIR for OpenMP/OpenACC, LLVM for OpenMP, NVIDIA for OpenMP, NVIDIA for OpenACC, GCC for OpenMP, GCC for OpenACC}

      \end{axis}
    \end{tikzpicture}

%% file: src/conclusion.tex
In this paper, we present UPIR, a unified parallel intermediate representation used for representing parallelism of parallel programming models to assist parallelism-aware compiler analysis, transformation and optimization. 
It is designed to support a wide-variety of parallel programming models and the prototype implementation in ROSE compiler support C/C++/Fortran, OpenMP, OpenACC and CUDA. 
UPIR enables a unified compiler transformation for multiple parallel programming models, including OpenMP and OpenACC.
Our experiments show that the UPIR compiler utilizes the unified transformation to compile both OpenMP and OpenACC programs.
It achieves promising performance and saves much development effort for supporting new programming models by leveraging the UPIR and the unified transformation. 
We believe that UPIR provides a comprehensive, flexible and extensible compiler IR designed for compiler development targeting modern heterogeneous parallel systems. Having the unified IR would enable lots of interesting research and accelerate the implementation of supporting new programming models in a compiler. 